\documentclass[twocolumn,showpacs,amsmath,amssymb,pra,floatfix,]{revtex4-1}

\usepackage{float}
\usepackage{epsfig}
\usepackage{amsmath}
\usepackage{amssymb}
\usepackage{bm}
\usepackage{graphicx}

\begin{document}

\title{A cosmic Zevatron based on cyclotron auto-resonance}

\author{Yousef I. Salamin}
\email{ysalamin@aus.edu}
\affiliation{Department of Physics and Materials Science and Engineering Research Institute, American University of Sharjah, POB 26666, Sharjah, United Arab Emirates}

\begin{abstract}
A Zevatron is an accelerator scheme envisaged to accelerate particles to ZeV energies (1 ZeV = $10^{21}$ eV). Zevatron schemes have been proposed to explain the acceleration of ultra-high-energy-cosmic-ray (UHECR) particles detected on Earth since 1962. It is shown in this Letter that nuclei of hydrogen, helium, and iron-56, may reach ZeV energies by cyclotron auto-resonance acceleration.   
\end{abstract}


\maketitle

\section{Introduction}

Cosmic rays are particles which reach Earth, with energies roughly in the range $10^8-10^{20}$ eV, and sometimes beyond, from deep space. As a result of 
collisions they make with the atmosphere, showers of secondary particles and flashes of light are produced, which can be seen by detectors on Earth \cite{honda,osmanov,linsley,auger}. The low-energy cosmic-ray particles are believed to originate in active stars 
and to gain their energies from the shock waves associated with such violent events as supernova explosions. It has also been suggested that particles which arrive with energies towards the end of the above range, and beyond, may be coming from active galactic nuclei (AGN) with massive black holes at their centers, or from the violent merger of neutron stars \cite{fang}, among other possibilities.  

Events of ultra-high-energy cosmic rays (UHECR) began to be detected more than 50 years ago \cite{linsley}, with energies greater than $10^{18}$ eV. It has been theoretically argued that the flux of UHECR particles gets attenuated as a result of interactions with the cosmic background radiation, so that only particles with energy below a maximum of about $4\times10^{19}$ eV, the so-called Greisen-Zatsepin-Kuz'min (GZK) limit \cite{greisen,zatsepin}, may reach Earth. However, in 1991 the first extreme-energy-cosmic-ray (EECR) particle was detected \cite{fly}, with kinetic energy exceeding $3\times10^{20}$ eV. Detection of UHECRs and EECRs meant that either the particles originated in places well within the radius dictated by the GZK limit, or that the limit itself had been violated by some unknown mechanism. 
Models have been proposed to explain both where the particles come from  and what mechanism of acceleration is responsible for the fantastic energies they carry \cite{honda,osmanov,fang,hillas}. This work is concerned with the energy issue, and aims to advance the scheme of cyclotron auto-resonance acceleration (CARA) as a possible resolution to it \cite{sal1,sal2,sal3,sal4,sal5}.

In CARA, a charged particle gains energy from the radiation field monotonically (by multi-photon absorption) if injected along the common direction of its propagation and that of a magnetic field, and provided an auto-resonance condition is also satisfied. Auto-resonance occurs when the cyclotron frequency of the particle, around the lines of the magnetic field, matches the Doppler-shifted frequency of the radiation field it senses. 

CARA has never been employed to explain acceleration of particles in extra-terrestrial environments, to the best of this author's knowledge. For it to work in accelerating nuclei to ZeV energies, as will be demonstrated below, the scheme requires the simultaneous presence, in some remote stellar environments, of mega- and giga-tesla magnetic fields, and the highest-power radiation fields believed to exist in the universe. Candidates for such environments include magnetars, merging neutron stars, and magnetar-powered supernova explosions. Radiation fields of the needed intensity may be associated with gamma-ray bursts.

The general problem of single-particle acceleration, in the simultaneous presence of a laser field and a uniform magnetic field, will be formulated and solved in Sec. \ref{sec:theory}. The main working equations, for auto-resonance acceleration, will be obtained in Sec. \ref{sec:equations}. Examples of auto-resonance acceleration of single protons and single nuclei of helium and iron will be investigated in Sec. \ref{sec:results}. A brief discussion of the work will be given in Sec. \ref{sec:discussion}.

\section{Theory}\label{sec:theory} 

Following is a brief review aimed at making this work self-contained \cite{sal1,sal2,sal3,sal4,sal5}. Consider a point particle, of mass $M$ and charge $+Q$, injected into a region of simultaneous presence of a uniform magnetic field and a linearly-polarized radiation field. The combined magnetic and radiation fields, in SI units, may be written, in a Cartesian coordinate system, as
\begin{equation}\label{EB}
\bm{E} = \hat{\bm{x}}E_0 \sin\eta;\quad \bm{B} = \hat{\bm{y}}\frac{E_0}{c} \sin\eta+\hat{\bm{z}}B_s,
\end{equation}
where $\hat{\bm{x}}, \hat{\bm{y}}$, and $\hat{\bm{z}}$ are unit vectors in the $x-$, $y-$, and $z-$directions, respectively, $E_0$ and $B_s$ are constants, and $\eta=\omega t-kz$ is the phase of the plane-wave radiation field, of frequency $\omega$ and wavevector $k$ (with $\omega = ck$).
The particle's relativistic momentum and energy are $\bm{p} = \gamma Mc\bm{\beta}$ and ${\cal E} = \gamma Mc^2$, respectively, where $\bm{\beta}$ is the velocity of the particle scaled by $c$, the speed of light in vacuum, and the Lorentz factor is $\gamma = (1-\beta^2)^{-1/2}$. 

The Newton-Lorentz equations of motion of the particle, in the above field combination, are
\begin{eqnarray}
\label{peq} \frac{d\bm{p}}{dt} &=& Q(\bm{E}+c\bm{\beta}\times\bm{B}),\\
\label{Eeq} \frac{d{\cal E}}{dt} &=& Qc\bm{\beta}\cdot\bm{\bm{E}}.
\end{eqnarray}
Subject to the rather simple initial conditions of position at the origin of coordinates and injection scaled velocity $\bm{\beta} = \beta_0\hat{\bm{z}}$, these equations admit closed-form analytic solutions, for the particle's trajectory and Lorentz factor \cite{sal1}. The steps leading to those solutions are fairly straightforward. First the $z-$ component of Eq. (\ref{peq}) and Eq. (\ref{Eeq}) read, respectively
\begin{eqnarray}
\label{pzeq} \frac{d}{dt} (\gamma\beta_z) &=& a_0\omega \beta_x\sin\eta;\quad a_0 = \frac{QE_0}{M\omega c},\\
\label{eeq} \frac{d\gamma}{dt} &=& a_0\omega \beta_x\sin\eta.
\end{eqnarray}
The initial conditions adopted above imply an initial value for the radiation field phase of $\eta_0=0$. With this in mind, the left-hand sides of these equations may be equated and the result integrated to yield a constant of the motion, namely
\begin{equation}\label{constant}
	\gamma(1-\beta_z) = \gamma_0(1-\beta_0); \quad \gamma_0=\frac{1}{\sqrt{1-\beta_0^2}}.
\end{equation}
The analytic solutions may best be arrived at if $\eta$ is employed to replace the time $t$ as a variable, with the following transformation playing a key role in the process
\begin{equation}\label{trans}
	\frac{d}{dt} = \omega(1-\beta_z)\frac{d}{d\eta}.
\end{equation}
In terms of $\eta$ and with the help of (\ref{constant}) the $x-$ and $y-$components of Eq. (\ref{peq}) now read, respectively
\begin{eqnarray}
\label{gbxe} \frac{d}{d\eta}(\gamma\beta_x) &=& a_0 \sin\eta+r(\gamma\beta_y),\\
\label{gbye} \frac{d(\gamma\beta_y)}{d\eta} &=& -r(\gamma\beta_x),
\end{eqnarray}
where
\begin{equation}\label{r}
	r = \frac{\omega_c}{\omega}\sqrt{\frac{1+\beta_0}{1-\beta_0}}; \quad \text{and}\quad \omega_c = \frac{QB_s}{M}.
\end{equation}
Note at this point that $a_0$ is a dimensionless radiation field strength, making $a_0^2$ a dimensionless intensity parameter. Also, $\omega_c$ is the cyclotron frequency of the particle around the lines of $\bm{B}_s$. This makes $r$ the ratio of the cyclotron frequency of the particle to the Doppler-shifted frequency of the radiation field, as sensed by the particle.

\begin{figure}[t]
	\centering
	\includegraphics[width=7.5cm]{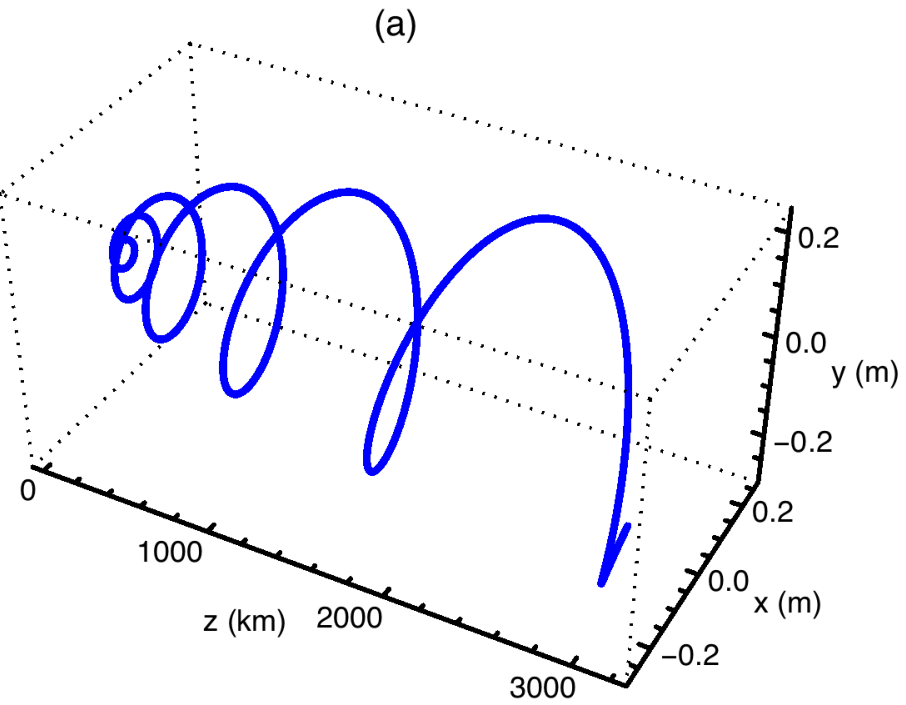}
	\includegraphics[width=8.5cm]{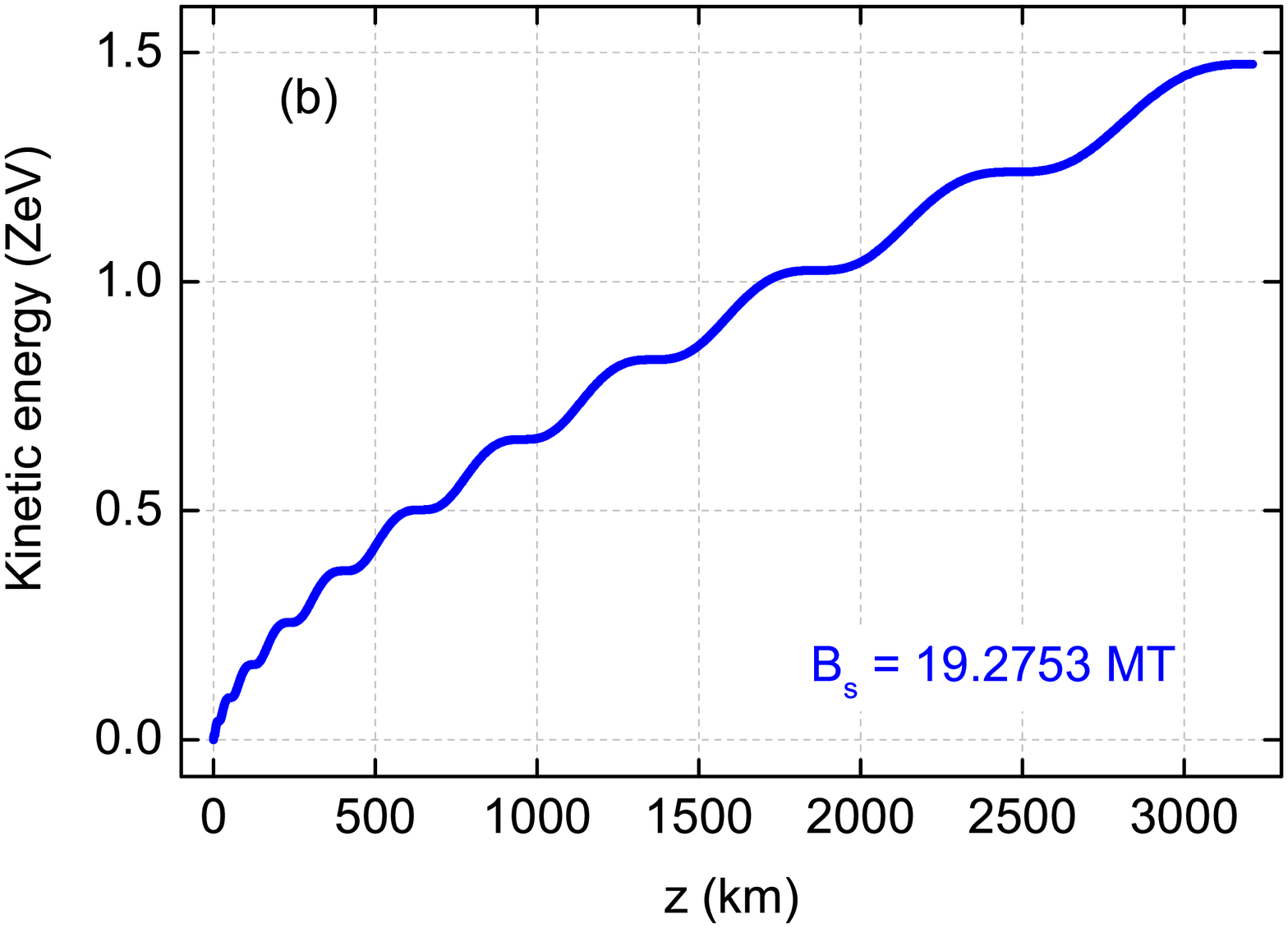}
	\caption{Example illustrating single-proton acceleration by cyclotron auto-resonance (CARA). (a) Trajectory, and (b) Kinetic energy, $K = (\gamma-1)Mc^2$, in the simultaneous presence of a uniform magnetic field of strength $B_s$, and a plane-wave, linearly-polarized, radiation field of wavelength $\lambda = 1~\mu$m and intensity $I = 4\times10^{38}$ W/m$^2$. The initial proton speed is $\beta_0 = 0.02$, and evolution is followed over $\Delta\eta = 12\pi$, or 6 field cycles.}
	\label{fig1}
\end{figure}

Subject to the same initial conditions, equations (\ref{gbxe}) and (\ref{gbye}) may be solved simultaneously, and yield
\begin{eqnarray}
	\label{gbx} \gamma\beta_x &=& a_0 \left[\frac{\cos \eta-\cos (r\eta )}{r^2-1}\right],\\
	\label{gby} \gamma\beta_y &=& a_0\left[\frac{\sin (r\eta )-r \sin \eta}{r^2-1}\right].
\end{eqnarray}
These equations give the $x-$ and $y-$ components of the particle's momentum, scaled by $Mc$. With the help of (\ref{constant}), (\ref{trans}) and (\ref{gbx}), Eq. (\ref{pzeq}) may be integrated, subject to same set of initial conditions, to give the following for the $z-$component of the scaled momentum
\begin{eqnarray}\label{gbz}
	\gamma\beta_z &=& \gamma_0\beta_0+a_0^2  \gamma_0(1+\beta_0) \left[\frac{\sin^2\eta}{2\left(r^2-1\right)}\right.\nonumber\\
	& & \left.+\frac{1-r \sin\eta \sin (r\eta)-\cos\eta \cos
   (r\eta)}{\left(r^2-1\right)^2}\right].
 \end{eqnarray}\label{gam}
Equation (\ref{gbz}) may be used to finally obtain the Lorentz factor of the particle from (\ref{constant})
\begin{equation}\label{gammaeq}
	\gamma = \gamma\beta_z+\gamma_0(1-\beta_0).
\end{equation}

An expression for the particle's $x-$coordinate, for example, as a function of $\eta$, may be obtained with the help of the transformation
\begin{equation}\label{dxdeta}
	\frac{dx}{d\eta} = \frac{dx/dt}{d\eta/dt}=\frac{c\beta_x}{\omega(1-\beta_z)}=\frac{c}{\omega}\gamma_0(1+\beta_0)(\gamma\beta_x).
\end{equation}
Using (\ref{gbx}) in (\ref{dxdeta}) and carrying out the integration over $\eta$, gives an expression for $x(\eta)$. Expressions for $y(\eta)$ and $z(\eta)$ may be obtained along the same lines. Finally, one gets the following parametric equations for the particle trajectory 
\begin{equation}\label{x}
	x(\eta) = \frac{ca_0}{\omega}\gamma_0(1+\beta_0) \left[\frac{r\sin\eta-\sin(r\eta)}{r(r^2-1)}\right], 
\end{equation}
\begin{widetext}
\begin{equation}\label{y}
	y(\eta) = \frac{ca_0}{\omega}\gamma_0(1+\beta_0) \left[\frac{1+r^2\cos\eta-\cos(r\eta)-r^2}{r(r^2-1)}\right],
\end{equation}
\begin{equation}\label{z}
	z(\eta) = \frac{c}{\omega}\left(\frac{1+\beta_0}{1-\beta_0}\right) \left\{\left[\frac{\beta_0}{1+\beta_0}
		+\frac{a_0^2}{4}\frac{3+r^2}{(r^2-1)^2}\right]\eta-\frac{a_0^2}{8}\frac{\sin(2\eta)}{(r^2-1)}+a_0^2\left[\frac{(1+r^2)\cos(r\eta)\sin\eta-2r\cos\eta\sin(r\eta)}{(r^2-1)^3}\right]\right\}.
\end{equation}
Furthermore, the Lorentz factor (\ref{gammaeq}) becomes
\begin{equation}\label{gamma}
	\gamma(\eta) = \gamma_0\left\{1+\frac{a_0^2}{2}(1+\beta_0)\left[\frac{[\cos(r\eta)-\cos\eta]^2 +[r\sin\eta-\sin(r\eta)]^2}{(r^2-1)^2}\right]\right\}.
\end{equation}
\end{widetext}

\section{The main working equations}\label{sec:equations}

In Eqs. (\ref{x})-(\ref{gamma}) the auto-resonance condition is identified as $r=1$. On resonance, and employing $\eta$ as a parameter, the solutions take on the following forms, obtained by taking the limits as $r\to1$ in (\ref{x})-(\ref{gamma}), respectively

\begin{equation}\label{xres}
x(\eta) = \frac{ca_0}{2\omega}\gamma_0(1+\beta_0) \left[\sin\eta-\eta\cos\eta\right],
\end{equation}
\begin{equation}
\label{yres}
y(\eta) = \frac{ca_0}{2\omega}\gamma_0(1+\beta_0) \left[\eta\sin\eta+2\cos\eta-2\right],
\end{equation}
\begin{eqnarray}
\label{zres}
z(\eta) &=& \frac{c}{\omega}\left(\frac{1+\beta_0}{1-\beta_0}\right) \left\{\left(\frac{\beta_0}{1+\beta_0}\right)\eta+\left(\frac{a_0^2}{24}\right)\eta^3\right.\nonumber\\
& &\left.
+\frac{a_0^2}{16}\left[2\eta\cos^2\eta-\sin(2\eta)\right]\right\},
\end{eqnarray}
\begin{equation}
\label{gammares}
\gamma(\eta) = \gamma_0\left\{1+\frac{a_0^2}{8}(1+\beta_0)\left[\eta^2+\sin^2\eta-\eta\sin(2\eta)\right]\right\}.
\end{equation}

\section{Results}\label{sec:results}

According to equation (\ref{gammares}) the Lorentz factor of an atomic nucleus, for example, increases during interaction with the radiation and magnetic fields, by a fraction $f = (\gamma-\gamma_0)/\gamma_0 \sim (Z/A)^2$, where $Z$ is the atomic number of the nuclear species, and $A$ is its mass number. This roughly means that an alpha particle and an iron nucleus can gain, respectively, 0.25 and $(30/56)^2$ times as much energy as a proton, when they interact with the same field environment, and provided the, admittedly delicate, auto-resonance condition is satisfied. Note, however, that the auto-resonance magnetic field strength is drastically different for different nuclear species, rendering the comparison entirely baseless.

In this environment, a particle gyrates around the lines of the magnetic field, and follows a helical trajectory of increasing radius, as Fig. \ref{fig1}(a) aims to demonstrate. For the parameters used in this example, the maximum radius of the helical trajectory is about 0.2 m, while the total axial excursion is over $3,200$ km. So, the particle's path is essentially a straight line. On the other hand, the particle's kinetic energy increases monotonically, by continuous multi-photon absorption from the radiation field. For the parameters employed, a proton's kinetic energy reaches 1.5 ZeV, as shown in Fig. \ref{fig1}(b). The trajectory and evolution of the kinetic energy are shown for interaction with 6 field cycles. The kinetic energy curve exhibits 12 {\it knees}. Each knee represents a {\it kick} in the particle's energy, due to interaction with one-half of a radiation field cycle. Figure 1(a) also shows a helical path of 6 windings. Each winding is the result of interaction with one complete cycle of the radiation field. There exists a one-to-one correspondence between the kicks in Fig. 1(b) and the windings of the helical trajectory (each winding corresponds to two successive kicks).

\begin{figure}[t]
	\centering
	\includegraphics[width=8.5cm]{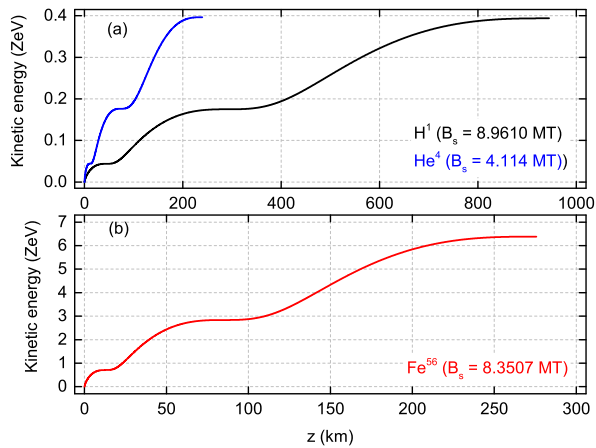}
	\caption{Kinetic energy evolution with axial excursion distance. results are shown for three nuclei, interacting with a uniform magnetic field of strength $B_s$, and a plane-wave, linearly-polarized (visible) radiation field of wavelength $\lambda = 1~\mu$m, and intensity $I = 4\times10^{38}$ W/m$^2$. For all particles, the initial injection speed is $\beta_0 = 0.9$, and evolution is followed over $\Delta\eta = 3\pi$. Every {\it knee} in each figure represents a jump in the energy gained, as a result of interaction with one-half of a radiation field cycle, or $\Delta\eta=\pi$.}
	\label{fig2}
\end{figure}

\begin{figure}[t]
	\centering
	\includegraphics[width=8.5cm]{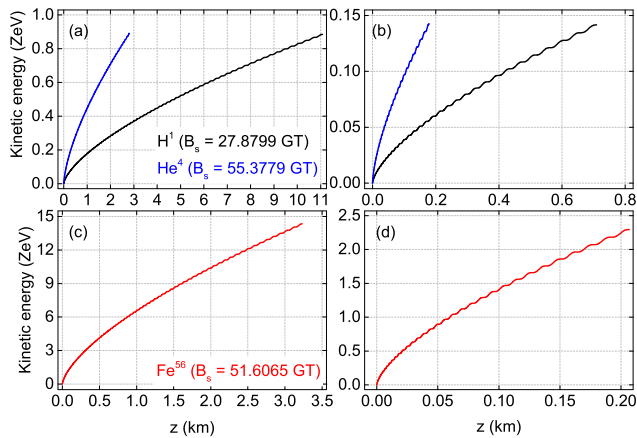}
	\caption{Kinetic energy evolution with axial excursion distance. Results are for three nuclei, during interaction with a magnetic field of strength $B_s$ and a gamma-ray burst of wavelength $\lambda=5\times10^{-11}$ m, and intensity $I=10^{44}$ W/m$^2$. For all particles, the injection speed $\beta_0=0.99$, and evolution is over $\Delta\eta=100\pi$ (50 field cycles). Each of (a) and (c) exhibits 100 hikes in the kinetic energy of the particle in question, with each hike the result of interaction with one-half of a radiation field cycle. (b) and (d) zoom, respectively, on small parts of (a) and (c).}
	\label{fig3}
\end{figure}

Figure \ref{fig1} is meant to illustrate CARA. The magnetic field strength needed to achieve auto-resonance in this particular example is $B_s \sim 1.92753\times10^7$ T, believed to be associated with {\it classical pulsars} \cite{reis}. In Figs. \ref{fig2} and \ref{fig3}, parameter values are employed, which are believed to be associated with magnetar-powered supernovae, according to recent studies \cite{greiner,sukhbold}. In such environments, magnetic fields of strengths $10^7-10^{11}$ T, and luminosities as high as $10^{46}$ J/s, seem to exist. Most of the energy released in such supernova explosions is radiant, carried away predominantly in the form of gamma-ray bursts, in addition to other radiation including visible. Figure \ref{fig2} is for super-intense visible light, and Fig. \ref{fig3} employs parameters akin to gamma-ray bursts. The figures suggest that UHECRs and EECRs are more likely to be iron nuclei, than protons. For protons to be accelerated by CARA and reach energies exceeding the GZK limit, after correction is made for the loss during the long journey from the source to Earth, the required mega- or giga-tesla magnetic fields need to be present over much longer distances than is currently believed to be in existence.

\section{Discussion}\label{sec:discussion}

A particle gains zero energy from interaction with an integer number of cycles of a plane-wave radiation field, in the absence of boundaries or a material medium, and under off-resonance conditions, according to the Lawson-Woodward theorem \cite{lw}. (One cycle corresponds to a change in phase $\Delta\eta=2\pi$.) This is because the particle gains energy during interaction with the positive (accelerating) half of a single cycle, only to lose it entirely during interaction with the following (identical) negative (decelerating) half. On resonance, however, the electric field vector of, say, a circularly-polarized plane wave, rotates about the direction of propagation at the frequency $\omega$. The particle, while in cyclotron motion about the lines of $\bm{B}_s$, senses the radiation field at the Doppler-shifted frequency $\omega_D=\omega\sqrt{(1-\beta_0)/(1+\beta_0)}$. The argument holds just as well for a linearly-polarized radiation field. Such a mode may be represented in terms of a superposition of two circularly-polarized ones, of opposite helicity. In either case, auto-resonance guarantees that the vectors $\bm{\beta}$ and $\bm{E}$ in Eq. ({\ref{eeq}) maintain the same orientation relative to each other. This means the rate $d{\cal E}/dt$ will remain positive, implying energy gain, during interaction with both halves of every radiation field cycle. Thus, energy of the particle will continue to increase monotonically.

Chances of the parameters $\beta_0$, $\omega$, $B_s$, and $Q/M$, conspiring to satisfy the auto-resonance condition are probably very small. Nevertheless, this should not come as a big surprise, in light of the fact that UHECR and EECR events are quite rare, indeed. For example, the Surface Array Experiment \cite{SAE} reported detecting 72 events, only, with energies of 57 EeV or more (1 EeV = $10^{18}$ eV) over the five-year period between 2008 and 2013.

\end{document}